# Long short-term memory networks in memristor crossbars

Can Li[1], Zhongrui Wang[1], Mingyi Rao[1], Daniel Belkin[1], Wenhao Song[1], Hao Jiang[1], Peng Yan[1], Yunning Li[1], Peng Lin[1], Miao Hu[2], Ning Ge[3], John Paul Strachan[2], Mark Barnell[4], Qing Wu[4], R. Stanley Williams[2], J. Joshua Yang[*1], Qiangfei Xia[*1]

[1] Department of Electrical and Computer Engineering, University of Massachusetts, Amherst, Massachusetts 01003, USA.
[2] Hewlett Packard Labs, HP Enterprise, Palo Alto, California 94304, USA
[3] HP Labs, HP Inc., Palo Alto, California 94304, USA.
[4] Air Force Research Laboratory, Information Directorate, Rome, New York 13441, USA

*Email: jjyang@umass.edu; qxia@umass.edu

## Abstract

Recent breakthroughs in recurrent deep neural networks with long short-term memory (LSTM) units has led to major advances in artificial intelligence. State-of-the-art LSTM models with significantly increased complexity and a large number of parameters, however, have a bottleneck in computing power resulting from limited memory capacity and data communication bandwidth. Here we demonstrate experimentally that LSTM can be implemented with a memristor crossbar, which has a small circuit footprint to store a large number of parameters and in-memory computing capability that circumvents the 'von Neumann bottleneck'. We illustrate the capability of our system by solving real-world problems in regression and classification, which shows that memristor LSTM is a promising low-power and low-latency hardware platform for edge inference.

**Introduction.**

The recent success of artificial intelligence largely results from the advances of deep neural networks with various microstructures\cite{lecun2015nature}, among which long short- term memory (LSTM) is an important unit\cite{lstm1997,lstm2000}. Enabling the learning process to remember or forget the history of observations, LSTM-based recurrent neural networks (RNNs) are responsible for recent achievements in analyzing temporal sequential data for applications such as data prediction \cite{gomez2005evolino, bao2017prediction}, natural language understanding \cite{jia2016data, karpathy2015unreasonable} machine translation \cite{google2016tranlate}, speech recognition \cite{microsoft2017speech}, and video surveillance \cite{lanz2017survillence}, etc. However, when implemented in conventional digital hardware, LSTM networks have complicated structures and hence drawbacks for inference latency and power consumption. These issues become more prominent as more applications involve the processing of temporal data near the source in the era of the Internet of Things (IoT). Although there has been an increased level of efforts in designing novel architectures to accelerate LSTMs based neural networks\cite{Chang2017, euge2017lstm_fpga, Cong2017lstm_fpga,ustc2017lstm_fpga, Conti2017, Gao2018DeltaRNN, Rizakis2018}low parallelism and limited bandwidth between computing and memory units are still outstanding issues. It is therefore imperative to seek an alternative computing paradigm for LSTM networks.

A memristor is a two-terminal 'memory resistor' \cite{chua1971memristor, strukov2008memristor}, which performs computation via physical laws at the same location where information is stored \cite{yang2013nnreview}. This feature removes the need for data transfer between memory and computation entirely. Built into a crossbar architecture, memristors have been successfully employed in feed-forward fully-connected neural networks

\cite{li2018analog, ibm2018mixed, strukov2015training, burr2015training, yu2016binary, wu2017face,hu2018dpe,li2018traininig} that showed significant advantages in power consumption and inference latency over CMOS-based counterparts \cite{xu2018scaling, cshwang2018review}. Short-term memory effects of some memristors were also utilized for reservoir computing \cite{lu2017reservoir}. On the other hand, most state-of-the-art deep neural networks are built with more sophisticated microstructures than fully-connected networks, in which LSTMs are responsible for the recent success of temporal data processing. The memristor crossbar implementation of an LSTM, to the best of our knowledge, has yet to be reported, primarily because of the relative scarcity of large memristor arrays.

In this work, we demonstrate our experimental implementation of LSTM networks in memristor crossbars. The memristors were monolithically integrated onto transistors forming one-transistor one-memristor (1T1R) cells. By connecting a memristor fully-connected network to a memristor recurrent LSTM network, we executed in-situ training and inference with the multilayer LSTM-based RNN for both regression and classification problems. The memristor LSTM network experiments succeeded in predicting airline passenger numbers and identifying an individual human based on gait. This work shows that the LSTM networks built in memristor crossbars represent a promising alternative computing paradigm with high speed-energy efficiency.

## Results

**Memristor crossbar for LSTM.** Neural networks containing LSTM units are recurrent, i.e. they not only fully connect the nodes in different layers, but also recurrently connect the nodes in the same layer at different time steps, as shown in Figure 1a. The recurrent connections in LSTM units also involve gated units to control the remembering and forgetting, which enable the learning of long-term dependencies \cite{lstm1997,lstm2000}. The data flow in a standard LSTM unit is shown in Figure 1b and is characterized by Equation 1 (linear matrix operations) and Equation 2 (gated nonlinear activations), or equivalently by Equations 3-5 in Methods.

$$\begin{bmatrix} \hat{\mathrm{a}}^t \\ \hat{\mathrm{i}}^t \\ \hat{\mathrm{f}}^t \\ \hat{\mathrm{o}}^t \end{bmatrix} = \begin{bmatrix} \mathrm{W_a} & \mathrm{U_a} & \mathrm{b_a} \\ \mathrm{W_i} & \mathrm{U_i} & \mathrm{b_i} \\ \mathrm{W_f} & \mathrm{U_f} & \mathrm{b_f} \\ \mathrm{W_o} & \mathrm{U_o} & \mathrm{b_o} \end{bmatrix} \begin{bmatrix} \mathrm{x}^t \\ \mathrm{h}^{t-1} \\ 1 \end{bmatrix} \tag{1}$$

$$\begin{aligned} \hat{\mathrm{c}}^t &= \sigma(\hat{\mathrm{i}}^t) \odot \tanh(\hat{\mathrm{a}}^t) + \sigma(\hat{\mathrm{f}}^t) \odot \hat{\mathrm{c}}^{t-1} \\ \mathrm{h}^t &= \sigma(\hat{\mathrm{o}}^t) \odot \tanh(\hat{\mathrm{c}}^t) \end{aligned} \tag{2}$$

where $\mathbf{x}^t$ is the input vector at the present step, $\mathbf{h}^t$ and $\mathbf{h}^{t-1}$ are the output vectors at the present and previous time steps respectively, $\hat{\mathbf{c}}^t$ is the internal cell state, and "$\odot$" is the element-wise multiplication. $\sigma$ is the logistic sigmoid function, which yields $\hat{\mathbf{i}}^t$, $\hat{\mathbf{f}}^t$, $\hat{\mathbf{o}}^t$ for the input, forget and output gates. The model parameters are stored in weights **W**, recurrent weights **U** and bias parameters **b** for cell activation (*a*) and each gate (*i, f, o*) respectively. Because of this complicated structure, state-of-the-art deep RNNs involving LSTM units include massive quantities of model parameters that typically exceeds the normal capacity of on-chip memory (usually static random access memory, SRAM), and sometimes even off-chip main memory (usually dynamic random access memory, DRAM). Consequently, the inference and training with the network will require the parameters to be transferred to the processing unit from a separate chip for the computation, and the data communication between chips heavily limits the performance of LSTM based RNNs on conventional hardware.

To address this issue, we have adopted a memristor crossbar for an RNN and store the large number of parameters required by an LSTM-RNN as the conductances of the memristors. The topography of this neural network architecture with the data flow direction is shown in Figure 1c. The linear matrix multiplications are performed *in situ* in a memristor crossbar, removing the need of transferring weight values back and forth. The model parameters are stored within the same memristor crossbar that performs the analog matrix multiplications. We connected an LSTM layer to a fully-connected layer for the experiments described here, and the layers can be cascaded into more complicated structures in the future. For demonstration purposes, the gated unit in the LSTM layer and the nonlinear unit in the fully-connected layer were implemented in software in the present work, but they can be implemented by analog circuits \cite{Smagulova2018memrsitor} without digital conversions to further significantly reduce the energy consumption and inference latency.

The analog matrix unit in our LSTM was implemented in a 128×64 1T1R crossbar with memristors monolithically integrated on top of a commercial foundry fabricated transistor array\cite{li2018analog} (Figures 2a-2c). The integrated Ta/$HfO_2$ memristors exhibited stable multilevel conductance that has enabled matrix multiplication in the analog domain \cite{Jiang2016Ta, li2018analog, hu2018dpe, li2018traininig}. With transistors controlling the compliance current, the integrated memristor array was programmed by loading a pre-defined conductance matrix with a write-and-verify approach (ex-situ training)\cite{ li2018analog, hu2018dpe} or by a simple two-pulse scheme in a fully-connected neural network (in-situ training)\cite{li2018traininig }. Inference in the LSTM layer was executed by applying voltages on the row wires of the memristor crossbar and reading out the electrical current through the virtual grounded column wires. The readout current vector is the dot product of the memristor conductance matrix with the input voltage-amplitude-vector, which was obtained directly by physical laws (Ohm's law for multiplication and Kirchhoff's current law for summation). Each parameter in the LSTM model was encoded by the conductance difference between two memristors in the same column, and subtraction was calculated in the crossbar by applying voltages with the same amplitude but different polarities on the corresponding row wires (Fig. 2a). The applied voltage amplitude on the rows that connect to the memristors for the bias representation is fixed across all the samples and time steps. The readout currents comprise four parts that represent the vectors $\hat{\mathbf{a}}^t, \hat{\mathbf{i}}^t, \hat{\mathbf{f}}^t$ and $\hat{\mathbf{o}}^t$ as described in Equation 1, which were nonlinearly activated and gated (Equation 2) and converted to voltages. The voltage vector ($\mathbf{h}^t$) was then fed into the next layer (a fully-connected layer in this work) and recurrently to the LSTM layer itself at the next time step ($\mathbf{h}^{t-1}$ at time $t$) (Figure 1c).

The neural network was trained in-situ within the memristor crossbar to compensate for possible hardware imperfection, such as limited device yield, variation and noise in conductance states \cite{yi2016noise_nc}, wire resistance, and analog peripheral asymmetry, etc. Before the training, all memristor conductances were initialized by one set voltage pulse across the memristor devices and synchronized gate voltages with a fixed amplitude. During the training, initial inferences were performed on a batch of sequential data (mini-batch) and yielded sequential outputs. After that, the memristor conductances were adjusted to make the inference outputs closer to the target outputs (evaluated by a loss function, see Methods). The intended conductance update values ($\Delta\mathbf{G}$) were calculated using the back-propagation through time (BPTT) algorithm\cite{sgdm, mozer1989bptt, werbos1988bptt} (see Methods for details). Th existing conductances in the memristor crossbar were updated with a two-pulse scheme that has previously been demonstrated to be effective in achieving linear and symmetric memristor conductance updates \cite{li2018traininig}.

**Regression experiment**. We first applied the memristor LSTM in predicting the number of airline passengers for the next month, a typical example of a regression problem. We built a two-layer RNN in a 128×64 1T1R memristor crossbar with each layer in a partition of the array. The input of the RNN was the number of air passengers in the present month, and the output was the projected number for the subsequent month. The RNN network structure is illustrated in Figure 3a. We used 15 LSTM units with a total of 2,040 memristors (34×60 array) representing 1,020 synaptic weights (Figure 3b), which took one data input, one fixed input for bias and 15 recurrent inputs from themselves. The second layer of the network was a fully-connected (FC) layer with 15 inputs from the LSTM layer and another input as the bias. The recurrent weights in the LSTM units represented the learned knowledge on when and what to remember and forget, and therefore the output of the network was dependent on both present and previous inputs.

The dataset we chose for this prediction task included the airline passenger number per month ranging from Jan 1949 to Dec 1960 with 144 observations\cite{airline_url}, in which the first 96 samples were selected as the training set, and the remaining 48 samples as the testing set (Figure 3c). During the inference, the number of passengers was linearly converted to a voltage amplitude (smaller than 0.2 V in order not to disturb the memristor conductances). The final output electrical current was scaled back to reflect the number of airline passengers. The training process is to minimized the mean square error (Equation 7) between the data in the training set and the network output, by the stochastic gradient descent through BPTT algorithm (see Methods). The raw voltages applied on the memristor crossbar and the raw output currents during the inference after 800 epochs are shown in Figures 3d-3g. The corresponding conductance and weight values are shown in Supplementary Figure 1, although they were not used for either inference or training process. The experimental training result in Figure 3c shows that the network learned to predict both the training data and the unseen testing data after 800 epochs of training.

**Classification experiment.** We further applied our memristor LSTM-RNN to identify an individual human by the person's gait. The gait as a biometric feature has a unique advantage when identifying a human from a distance, as other biometrics (e.g. face) occupy too few pixels to be recognizable. It becomes increasingly important in circumstances in which face recognition is not feasible because of camouflage and/or lack of illumination. For employing gait in a surveillance application scenario, it is preferable to deploy many cameras and perform the inference locally rather than sending the raw video data back to a server in the cloud. Inference near the source should be performed with low-power and small communication bandwidth, but still achieve low-latency.

The memristor LSTM-RNN utilized a feature vector extracted from a video frame as the input, and outputs the classification result as electrical current at the end of the sequence. (Figure 4a). We implemented the two-layer RNN by partitioning a 128×64 memristor crossbar (Figure 4b), in which 14 LSTM units in the first layer were fully connect to the 50-dimensional input vector with 64×56 connections (implemented in a 128×56 memristor crossbar). The 14 LSTM units further fully connected to eight output nodes. The classification result was represented by the maximum dimension in the output vectors of the output nodes in the fully-connected layer.

To demonstrate the core operation of the memristor LSTM memristor network, the feature vectors for the input of the LSTM-RNN were extracted from video frames by software. Human silhouettes with 128×88 pixels were first pre-extracted from the raw video frames in the USF-NIST gait dataset \cite{usf2002gait} and then processed into 128-dimensional width-profile-vectors \cite{kale2004widthvector}. The vectors were then down-sampled to 50 dimensions to fit the size of our crossbar (Figure 4c). We chose the video sequences from eight different people out of 75 in the original dataset. The videos cover various scenarios with people wearing two different pairs of shoes on two different surface types (grass or concrete) taken from two different viewpoints (eight co-variance). The videos sequences were further segmented into 664 sequences each with 25 frames, as described in detail in Supplementary Figure 2. The training was performed on 597 sequences randomly drawn from the dataset, while the remaining 67 unseen sequences were used for the classification test. In state-of-the-art deep neural networks, the feature vectors that feed into the LSTM layer are usually extracted by multiple convolutional layers and/or fully connected layers without much human knowledge. The feature extraction step could also be implemented in a memristor crossbar when multiple arrays are available in the near future \cite{li2018analog}.

The training and inference processes were experimentally performed in the memristor crossbar, with a procedure similar to that in the regression experiment (see Methods). The goal of the training was to minimize the cross-entropy in the Bayesian probability (Equation 8 in Methods), which was the loss function, that is calculated from the last time step electrical current and the ground truth (Figure 4a). The desired weight update values were optimized with root mean square propagation (RMSprop) \cite{rmsprop} based on the calculated weight gradient by the BPTT algorithm, and applied to the memristor crossbar after the inference operation on one mini-batch of 50 training sequences. The mean cross-entropy during the inference of each mini-batch was calculated and shown in Figure 4d, from which one sees the effectiveness of the training with the memristor crossbar. The classification test was conducted on the separate testing set after training on each epoch. The classification accuracy increased steadily during the training, and the maximum accuracy within the 50 epochs of training was 79.1%, which closely matched the defect-free simulation (Figure 4e), confirming that the in-situ training adapted to the hardware imperfections without hand-tuned parameters.

## Discussion

In summary, we have built multilayer RNNs with a memristor LSTM layer and a memristor fully-connected layer. The successful demonstrations on both regression and classification tasks exhibited the versatility of connecting the memristor neural network layers with different configurations. The results open up a new direction for integrating multiple memristor crossbars with different configurations on the same chip, which will minimize data transfer and significantly reduce the inference latency and power consumption in a deep recurrent neural network.

## Methods

**1T1R array integration.** The transistor array was fabricated in a commercial foundry using the 2 μm technology node. We then integrated our memristors in a university cleanroom. The transistors were used as selector devices to mitigate the sneak path problem in the crossbar and to enable precise conductance tuning. Two layers of metal wires were also fabricated in the foundry back-end-of-the-line (BEOL) process as row and column wires to reduce the wire resistance (about 0.3 Ω between cells). The low wire resistance in the array is one of the key factors that provided accurate matrix multiplication. The memristors were fabricated on top of the transistor array in the UMass Amherst cleanroom, with sputtered palladium as the bottom electrode, atomic layer deposited hafnia as the switching layer and sputtered tantalum as the top electrode.

**Inference in the two layer LSTM-RNN.** The network in this work had two layers, with the first layer being the LSTM and the second a fully-connected layer. The algorithm can extend to more layers because of the cascaded structure. The input to the network was $\mathbf{x}^t$, and the LSTM cell activation $\mathbf{a}^t$ was calculated as shown EQUATION 3.

$$\mathbf{a}^t = \tanh(\hat{\mathbf{a}}^t) = \tanh\left(\mathbf{W}_{\mathrm{a}}\mathbf{x}^t + \mathbf{U}_{\mathrm{a}}\mathbf{h}^{t-1} + \mathbf{b}_{\mathrm{a}}\right) \quad (3)$$

The input gate, forget gate and the output gate that control the output are defined in EQUATION 4

$$\begin{aligned} \mathbf{i}^t &= \sigma(\hat{\mathbf{i}}^t) = \sigma(\mathbf{W}_{\mathrm{i}}\mathbf{x}^t + \mathbf{U}_{\mathrm{i}}\mathbf{h}^{t-1} + \mathbf{b}_{\mathrm{i}}) \\ \mathbf{f}^t &= \sigma(\hat{\mathbf{f}}^t) = \sigma(\mathbf{W}_{\mathrm{f}}\mathbf{x}^t + \mathbf{U}_{\mathrm{f}}\mathbf{h}^{t-1} + \mathbf{b}_{\mathrm{f}}) \\ \mathbf{o}^t &= \sigma(\hat{\mathbf{o}}^t) = \sigma(\mathbf{W}_{\mathrm{o}}\mathbf{x}^t + \mathbf{U}_{\mathrm{o}}\mathbf{h}^{t-1} + \mathbf{b}_{\mathrm{o}}) \end{aligned} \quad (4)$$

The output of the LSTM layer (as the hidden layer output in the two-layer RNN) was determined by EQUATION 5. EQUATION 3, 4, 5 are equivalent to EQUATION 1, 2 in the main text, in which the linear and nonlinear operations are separated for easier comprehension.

$$\begin{aligned} \mathbf{c}^t &= \tanh\left(\mathbf{i}^t \odot \mathbf{a}^t + \mathbf{f}^t \odot \hat{\mathbf{c}}^{t-1}\right) \\ \mathbf{h}^t &= \mathbf{o}^t \odot \mathbf{c}^t \end{aligned} \quad (5)$$

The final output of the RNN was read out by a fully-connected layer, and the function of which is characterized by EQUATION 6

$$\mathbf{y}^t = f\left(\hat{\mathbf{y}}^t\right) = f\left(\mathbf{W}_{\mathrm{FC}}\mathbf{h}^t + \mathbf{b}_{\mathrm{FC}}\right) \quad (6)$$

where $f$ is the nonlinear activation function in the fully-connected layer. Specifically, we used the logistic sigmoid function in the airline prediction experiment and the softmax function in the human gait identification experiment.

**Training with back-propagation through time (BPTT)** The goal of the training process was to minimize a loss function, which was a function of the network output $\mathbf{y}^t$ and their targets $\mathbf{y}^t$ (ground truth or labels). Specifically, we chose mean square loss error over all time steps for the airline prediction experiment (EQUATION 7) and cross-entropy loss on the last time step for the human gait identification experiment (EQUATION 8).

$$\mathcal{L}_{\text{airline}} = \sum_{n=1}^{N}\sum_{t=1}^{T}\frac{1}{2}\left(\mathbf{y}^{t}(n) - \mathbf{y}^{t}_{\text{target}}(n)\right)^{\top}\left(\mathbf{y}^{t}(n) - \mathbf{y}^{t}_{\text{target}}(n)\right)/T \tag{7}$$

$$\mathcal{L}_{\text{gait}} = -\sum_{n=1}^{N}\sum_{c=1}^{C} y_{c,\text{target}}(n)\cdot\log\left[y_c^{T}(n)\right] \tag{8}$$

where $n$ indexes over the sample, $N$ is the batch size, $t$ is the temporal sequence number, and $T$ is the total of the time steps in the sequence.

The training, i.e. model optimization, was based on the weight gradients of the loss function. Since the weights stayed the same in the same mini-batch over all the time steps, the gradients were accumulated before each weight update. The gradient of the loss function $L$ on sample $n$ at sequence $t$ is denoted as $\delta\mathbf{v}^{t} = \frac{\partial L}{\partial \mathbf{v}^{t}}$ , and is calculated by the backpropagation through time (BPTT) algorithm \cite{mozer1989bptt, werbos1988bptt}. The last layer output delta was calculated by EQUATION 9 for the airline prediction task and by EQUATION 10 for the gait identification task.

$$\delta\hat{\mathbf{y}}^{t} = \frac{\partial\mathcal{L}}{\partial\hat{\mathbf{y}}^{t}} = \sigma'(\mathbf{y}^{t} - \mathbf{y}^{t}_{\text{target}}) \tag{9}$$

where $\sigma'$ is the derivative of the logistic sigmoid function.

$$\delta\hat{\mathbf{y}}^{t} = \frac{\partial\mathcal{L}}{\partial\hat{\mathbf{y}}^{t}} = \begin{cases}\mathbf{y}^{t} - \mathbf{y}^{t}_{\text{target}} & t = T;\\ 0 & t < T.\end{cases} \tag{10}$$

The previous layer deltas were calculated with the chain rule.

$$\delta \mathbf{h}^t = \mathbf{W}_{\text{FC}}^\top \delta \hat{\mathbf{y}}^t + \delta \mathbf{h}^{t+1} \tag{11}$$

$$\delta \hat{\mathbf{o}}^t = \delta \mathbf{h}^t \odot \mathbf{c}^t \odot \sigma'(\mathbf{o}^t) \tag{12}$$

$$\delta \hat{\mathbf{c}}^t = \delta \mathbf{h}^t \odot \mathbf{o}^t \odot \tanh'(\mathbf{c}^t) + \delta \hat{\mathbf{c}}^{t+1} \tag{13}$$

$$\delta \hat{\mathbf{a}}^t = \delta \hat{\mathbf{c}}^t \odot \mathbf{i}^t \odot \tanh'(\mathbf{a}^t) \tag{14}$$

$$\delta \hat{\mathbf{i}}^t = \delta \hat{\mathbf{c}}^t \odot \mathbf{a}^t \odot \sigma'(\mathbf{i}^t) \tag{15}$$

$$\delta \hat{\mathbf{f}}^t = \delta \hat{\mathbf{c}}^t \odot \hat{\mathbf{c}}^{t-1} \odot \sigma'(\mathbf{f}^t) \tag{16}$$

$$\delta \hat{\mathbf{c}}^{t-1} = \delta \hat{\mathbf{c}}^t \odot \hat{\mathbf{f}}^t \tag{17}$$

$$\begin{bmatrix} \delta \mathbf{x}^t \\ \delta \mathbf{h}^{t-1} \end{bmatrix} = \begin{bmatrix} \mathbf{W_a}^\top & \mathbf{W_i}^\top & \mathbf{W_f}^\top & \mathbf{W_o}^\top \\ \mathbf{U_a}^\top & \mathbf{U_i}^\top & \mathbf{U_f}^\top & \mathbf{U_o}^\top \end{bmatrix} \begin{bmatrix} \delta \hat{\mathbf{a}}^t \\ \delta \hat{\mathbf{i}}^t \\ \delta \hat{\mathbf{f}}^t \\ \delta \hat{\mathbf{o}}^t \end{bmatrix} \tag{18}$$

The computationally expensive steps described in EQUATION 11 and EQUATION 18 were calculated in the crossbar. The weight gradients were calculated based on the delta rule.

$$\begin{bmatrix} \delta \mathbf{W}_{\text{FC}}^t & \delta \mathbf{b}_{\text{FC}}^t \end{bmatrix} = \begin{bmatrix} \delta \hat{\mathbf{y}}^t \end{bmatrix} \begin{bmatrix} \mathbf{h}^{t\top} & 1 \end{bmatrix} \tag{19}$$

$$\begin{bmatrix} \delta \mathbf{W_a}^t & \delta \mathbf{U_a}^t & \delta \mathbf{b_a}^t \\ \delta \mathbf{W_i}^t & \delta \mathbf{U_i}^t & \delta \mathbf{b_i}^t \\ \delta \mathbf{W_f}^t & \delta \mathbf{U_f}^t & \delta \mathbf{b_f}^t \\ \delta \mathbf{W_o}^t & \delta \mathbf{U_o}^t & \delta \mathbf{b_o}^t \end{bmatrix} = \begin{bmatrix} \delta \hat{\mathbf{a}}^t \\ \delta \hat{\mathbf{i}}^t \\ \delta \hat{\mathbf{f}}^t \\ \delta \hat{\mathbf{o}}^t \end{bmatrix} \begin{bmatrix} \mathbf{x}^{t\top} & \mathbf{h}^{t-1\top} & 1 \end{bmatrix} \tag{20}$$

The parameters (weights or bias) gradients were accumulated as described in EQUATION 21.

$$\mathbf{GRAD} = \sum_{t=1}^{T} \sum_{n=1}^{N} \delta \mathbf{W}^{t}(n) \tag{21}$$

The stochastic gradient descent with momentum (SGDM) optimizer that we used in the airline prediction problem yielded the desired weight update value by EQUATION 22.

$$\mathbf{\Delta W} = -\left(\eta \mathbf{\Delta W}_{\text{pre}} + \alpha \mathbf{GRAD}\right) \tag{22}$$

where $\eta$ and $\alpha$ are the hyper-parameters for momentum and learning rate, respectively.

In the gait identification experiment, we used the root mean square propagation (RMSprop) optimizer, which gives the desired weight update values by EQUATION 23.

$$\mathbf{MS} = \beta \mathbf{MS}_{\text{pre}} + (1 - \beta) \mathbf{GRAD}^{\circ 2} \tag{23}$$

$$\Delta \mathbf{W} = -\left[\eta \mathbf{\Delta W}_{\text{pre}} + \alpha \mathbf{GRAD} \oslash \left(\sqrt{\mathbf{MS}} + \epsilon\right)\right] \tag{24}$$

where $\beta$, $E$, $\alpha$ and $\eta$ are hyper-parameters, $\mathbf{GRAD}^{\circ 2}$ is the element-wise square operation on matrix **GRAD** and $\oslash$ indicates the element-wise division operation.

**Hyperparameters.** The following table shows the hyperparameters during the training experiment. They include both the hyperparameters for the neural network, and the physical parameters to operate the memristor crossbar.

**Table 1:** Training hyperparameters and physical parameters of the memristor crossbar for the regression and classification experiments

| Parameter | Airline prediction | Gait identification | Description |
|---|---|---|---|
| $\alpha$ | 0.01 | 0.01 | Learning rate |
| $\eta$ | 0.9 | 0 | Momentum |
| $\beta$ | N/A | 0.9 | Decay for RMSprop |
| $\varepsilon$ | N/A | $1 \times 10^{-8}$ | Denominator shift for RMSprop |
| $G/W$ | $1 \times 10^{-4}$ | $3 \times 10^{-4}$ | Conductance-to-weight ratio |
| $\Delta V_{gate}/\Delta G$ | $1.02 \times 10^{4}$ | $1.02 \times 10^{4}$ | Gate-voltage-to-conductance ratio |
| $V_{gate,\ 0}$ | 1.0 | 1.0 | Initial gate voltage |
| $V_{set}$ | 2.5 | 2.5 | Set voltage |
| $V_{reset}$ | 1.7 | 1.7 | Reset voltage |
| $V_{read}$ | 0.2 | 0.2 | Read voltage |
| $V_{gate,\ max}$ | 1.6 | 1.6 | Maximum set gate voltage |
| $V_{gate,\ min}$ | 0.7 | 0.7 | Minimum set gate voltage |
| $V_{gate,\ reset}$ | 5.0 | 5.0 | Reset gate voltage |

**Data availability.** The data that support the plots within this paper and other finding of this study are available from the corresponding author upon reasonable request.

**Acknowledgements.** This work was supported in part by the U.S. Air Force Research Laboratory (AFRL; grant no. FA8750-15-2-0044) and the Intelligence Advanced Research Projects Activity (IARPA; contract no. 2014-14080800008). Daniel Belkin, an undergraduate from Swarthmore College was supported by NSF research experience for undergraduates (REU; grant no. ECCS-1253073) at UMass. Peng Yan was visiting from Huazhong University of Science and Technology under the support from the Chinese Scholarship Council (CSC) (grant 201606160074). Part of the device fabrication was conducted in the clean room of Center for Hierarchical Manufacturing (CHM), an NSF Nanoscale Science and Engineering Center (NSEC) located at the University of Massachusetts Amherst.

**Author contributions.** Q.X. conceived the idea. Q.X., J.J.Y., C.L. designed the experiments. C.L., Z.W., D.B. did the programming, measurements, data analysis and simulation. M.R., P.Y, C.L., H.J., N.G., P.L. built the integrated chips. Y.L., E.M., C.L., W.S., M.H., Z.W., J.P.S. built the measurement system and firmware. Q.X., C.L., J.J.Y., R.S.W. wrote the manuscript. M.B., Q.W. and all other authors contributed to the result analysis and commented on the manuscript.

**Competing Interests.** The authors declare that they have no competing financial or non-financial interests.

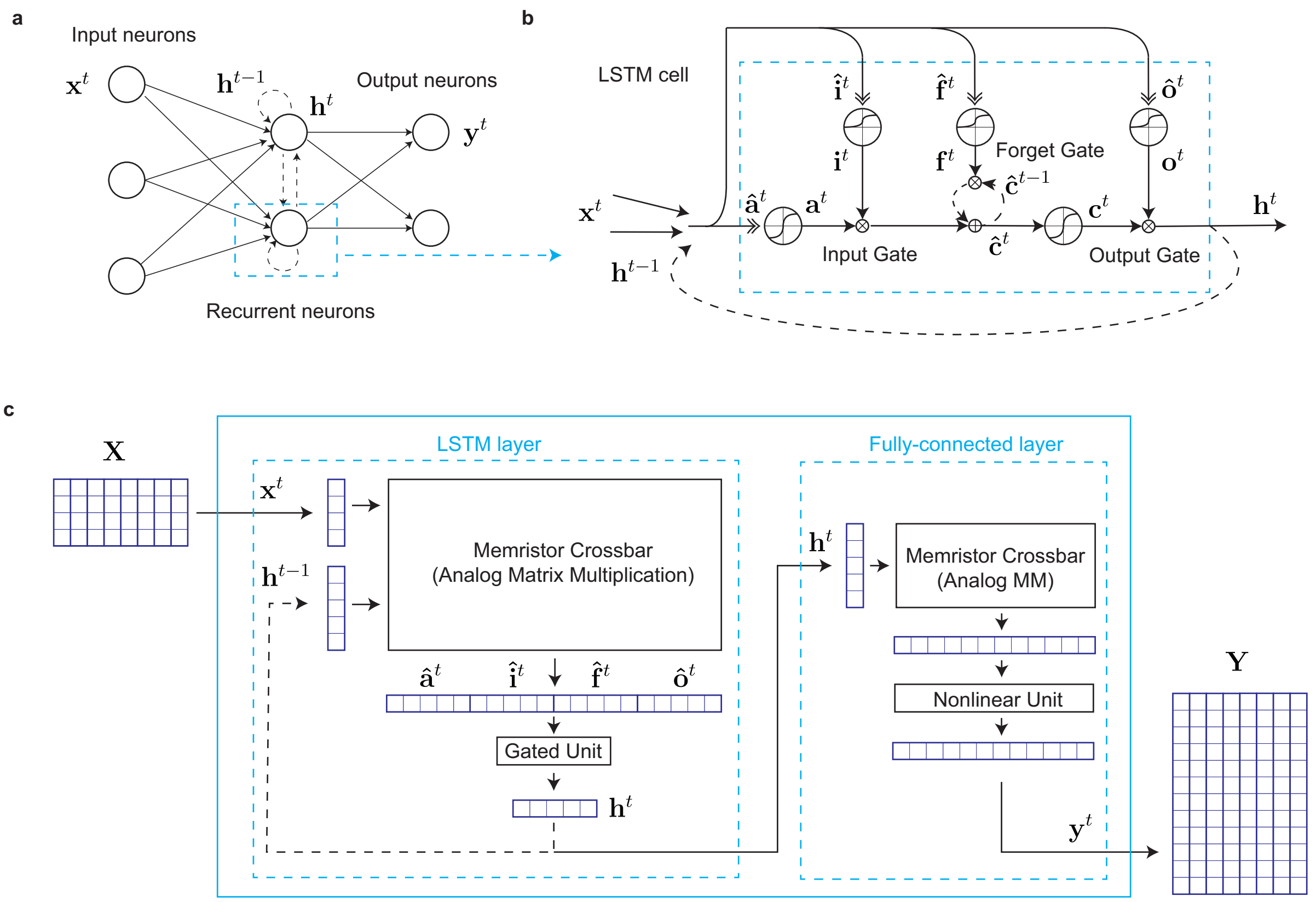

**Figure 1: Schematic architecture of memristor accelerated long short-term memory (LSTM) network. a,** Schematic of a multilayer recurrent neural network (RNN) with input nodes, recurrent hidden nodes and output nodes. The recurrent nodes (e.g. LSTM units) fully connect to both the input nodes and the previous state of the recurrent nodes. **b,** The structure of a standard LSTM cell (blue dashed box in **a**), which includes input, forget, and output gates to resolve the vanishing or exploding gradient problems in standard RNN units and learn long-term dependencies. **c,** The data flow in the present architecture. Input and output (I/O) data, X and Y, are sent to/from the integrated chip (blue box) through off-chip peripheral circuits. The figure shows a two-layer RNN which is composed of a LSTM layer and a fully-connected layer. For both layers, the synaptic connections (parameters) are stored *in-situ* in the crossbar as conductances, minimizing the data communication. In this work, the gated unit in the LSTMs and the nonlinear unit in the fully connected layers were implemented in software. The LSTM layer and fully connected layer can be cascaded to produce more hardware layers in hardware in future work.

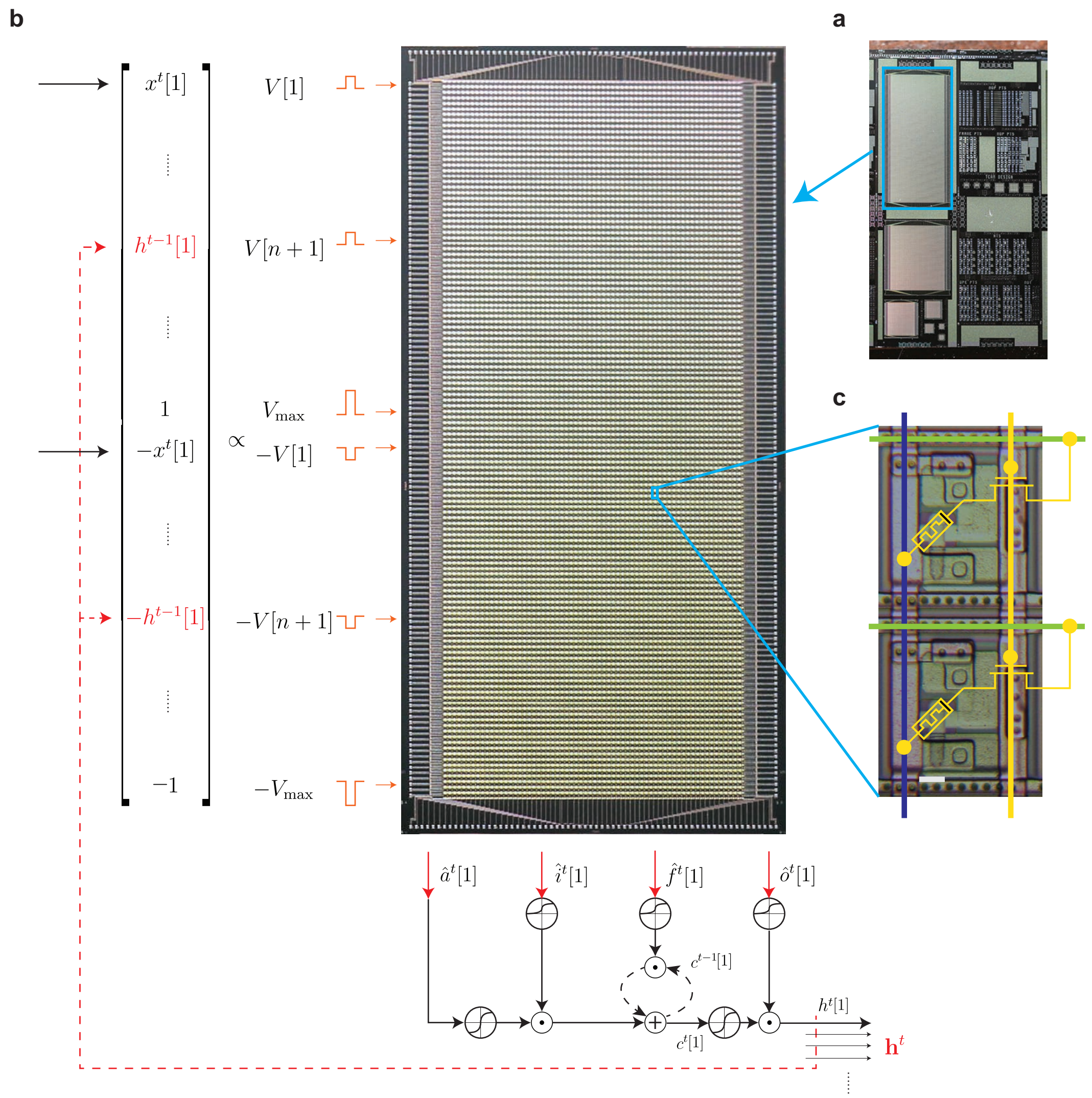

**Figure 2: LSTM units implemented in a memristor crossbar. a,** One die one the integrated 1T1R chip with various array sizes (from 4×4 to 128×64) and testing circuits. **b,** Part of the 128×64 one-transistor one-memristor (1T1R) integrated array is used for the LSTM. The input vectors (**x**) are converted to small analog voltages (V), which are applied on the row wires of the memristor crossbar while the column wires are grounded. The currents in different columns, which represent the solutions of matrix-vector multiplication, are labelled as $\hat{\mathbf{a}}^t$, $\hat{\mathbf{i}}^t$, $\hat{\mathbf{f}}^t$, $\hat{\mathbf{o}}^t$, respectively. The current vectors are then nonlinearly activated and gated to yield the LSTM output vector $\mathbf{h}^t$, which is fed an inputs to the next layer, and the present layer of the next time step. Scale bar, 500 µm **c,** Enlarged images of two 1T1R cells, with a circuit diagrams that show the electrical connections. During the inference, voltages are applied on the row wires (green), and currents are read from the column wires (blue). High voltages (~5 V) are applied on the gate wires to turn on all the transistors. Scale bar, 10 µm.

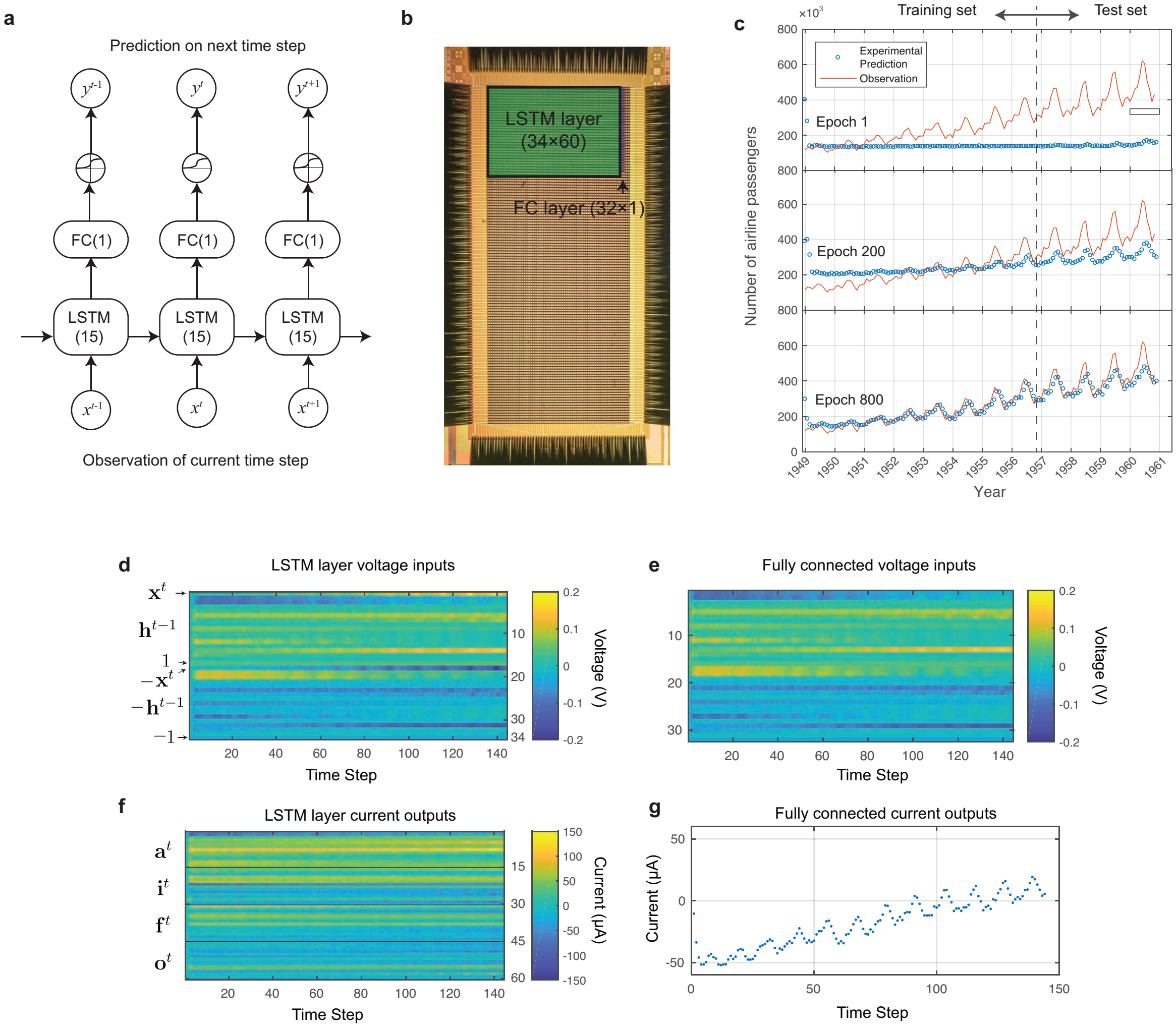

**Figure 3: Regression experiment for predicting the next month's number of airline passengers. a**, Architecture of the two-layer recurrent neural network (RNN) configured for prediction. The input $x^t$ is the observed number of passengers month $t$, and the output $y^t$ is the predicted number for the month $t + 1$. We used 15 LSTM units with 2,040 memristors to represent the required 1,020 synaptic weights. The output node was fully connected to the 15 LSTM output nodes by 32 memristors, and the final prediction was the nonlinear activation of the fully-connected (FC) layer output after filtering by the logistic sigmoid function. **b**, Partition of the 128×64 1T1R memristor array. A 34×60 sub-array was used for the LSTM layer and a 32×1 sub-array was used for the FC layer. **c**, The in-situ training and test results of the two-layer RNN. Two-thirds of the data are used for training, while the remaining was used as the test set. The network precisely predicted the future airline passenger numbers after training for 800 epochs. **d-g**, Raw voltage inputs and electrical current outputs for the LSTM layer (d, f) and fully-connected layer (e, g). The data presented in (c) is after activation and scaling of the data in (g).

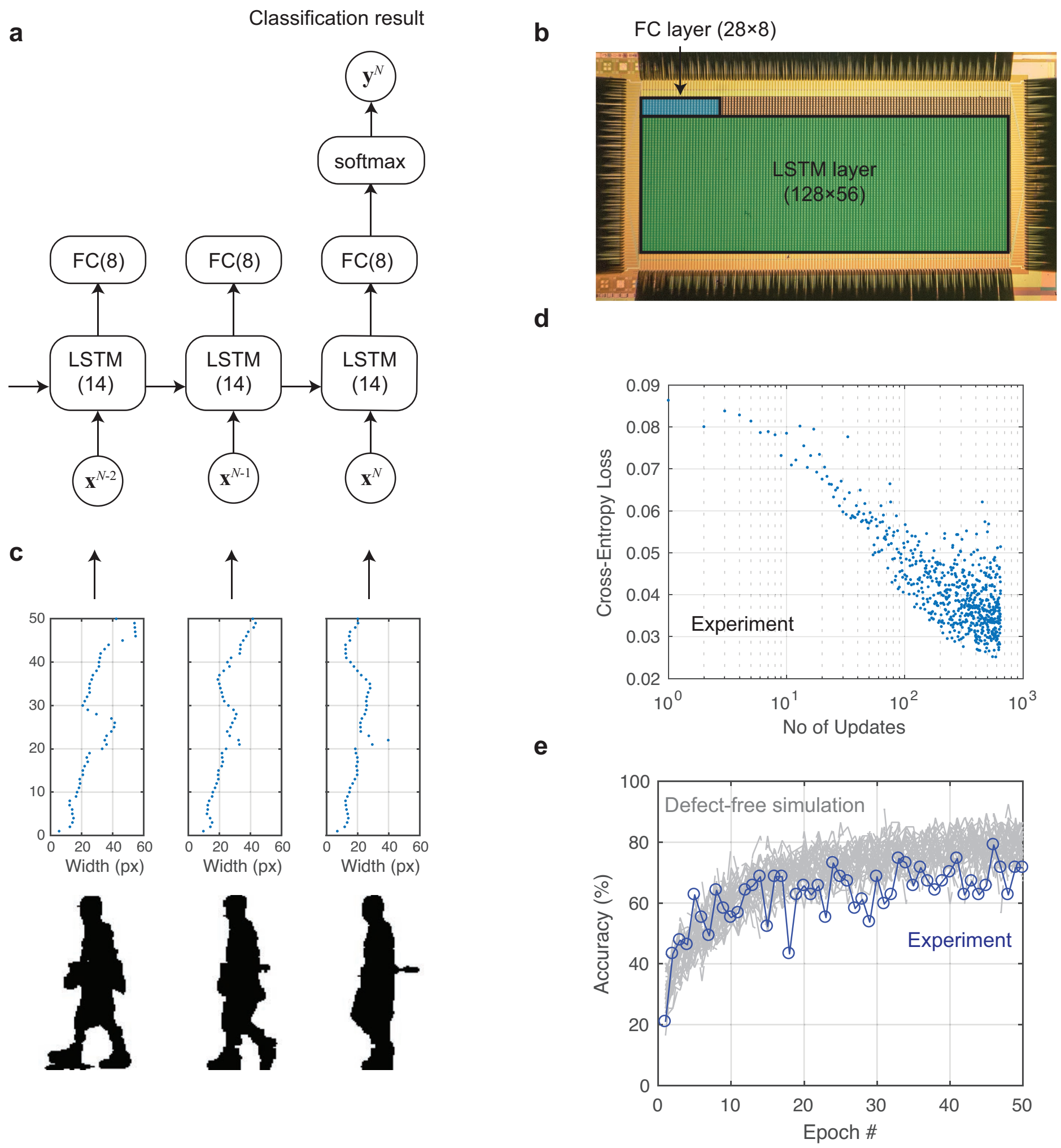

**Figure 4: Classification experiment for human identification by its gait. a,** The two-layer recurrent neural network (RNN) configuration for classification. In the RNN, there were 14 LSTM units, with 7,168 memristors representing the synaptic weights. The eight output nodes were fully connected with the 14 LSTM nodes by 228 memristors, and a person was identified by the maximum electrical current output in the last time step. The training of the network is to minimized the cross-entropy between the predicted softmax probability in the final step ($\mathbf{y}^N$) and ground truth, where $N$ was the length of the temporal sequence. **b,** Partition of the 128×64 1T1R memristor crossbar, in which a 128×56 sub-array was used for the LSTM layer and a 28×8 sub-array was used for the fully-connected layer. **c,** We used the width profiles of the human silhouettes extracted from a video as the inputs for the RNN. The pre-processing of the images is illustrated in SUPPLEMENTARY FIGURE 2. **d,** The cross-entropy loss steadily decreases during the training, showing the effectiveness of training of RNN with LSTM units. **e,** Classification accuracy on the unseen testing set during the 50 epochs of in-situ training. The gray lines are the results from 50 repeated simulations of a defect-free crossbar with randomly initialized weights. Experimentally (blue line), we approached a maximum of 79.1% accuracy, which closely matched the defect-free simulation, showing that the training step included significant defect tolerance. The experimental conductance and weight values awerere readout as shown in SUPPLEMENTARY FIGURE 3, and the raw current output and calculated Bayesian probability in the classification test after training is shown in SUPPLEMENTARY FIGURE 4 for reference.

## *Supplementary Figures*

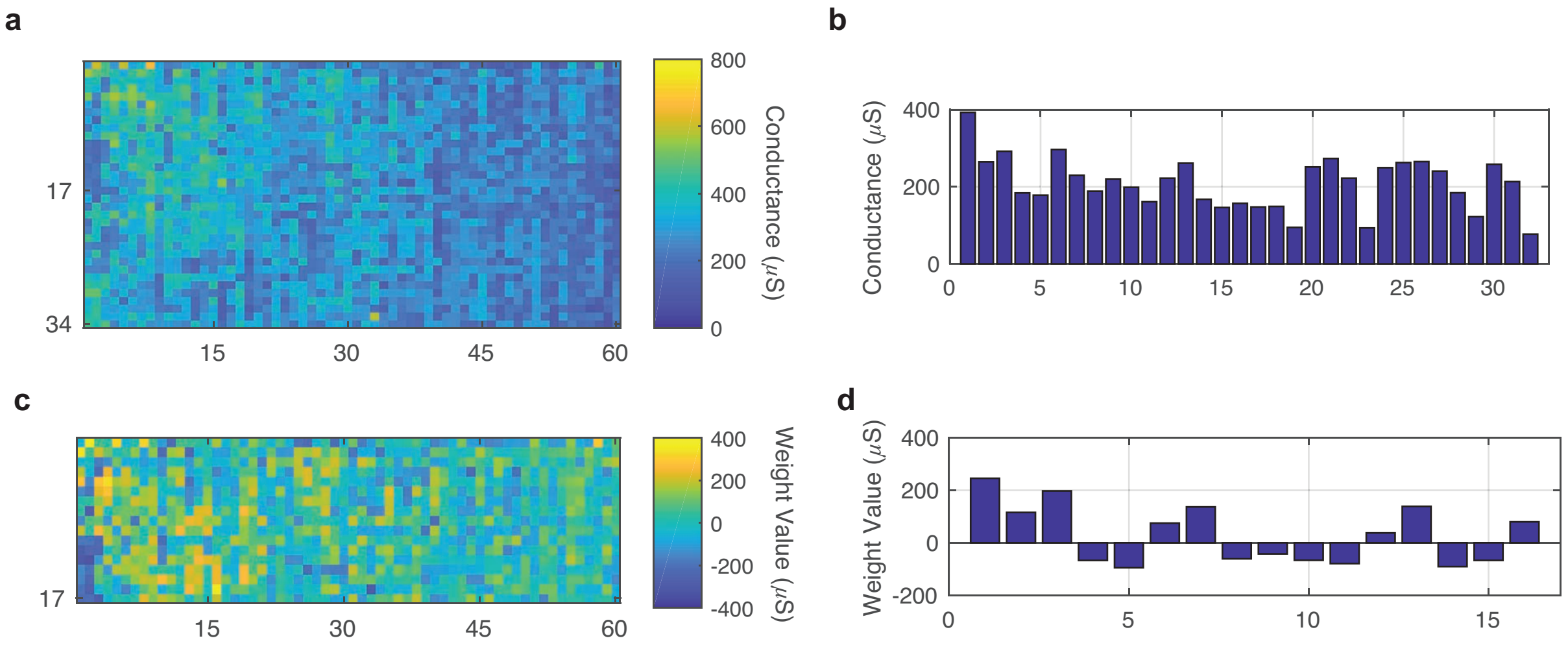

**Supplementary Figure 1: Additional data for the regression experiment. a,** Conductance map of the 34×60 memristor array in the LSTM layer after the in-situ training. **b,** Measured conductance of the 32 memristors in the fully-connected layer after training. **c,** Map of synaptic weight calculated from the conductances shown in *(a)*. **d,** Synaptic weights calculated from the conductances shown in *(b)*.

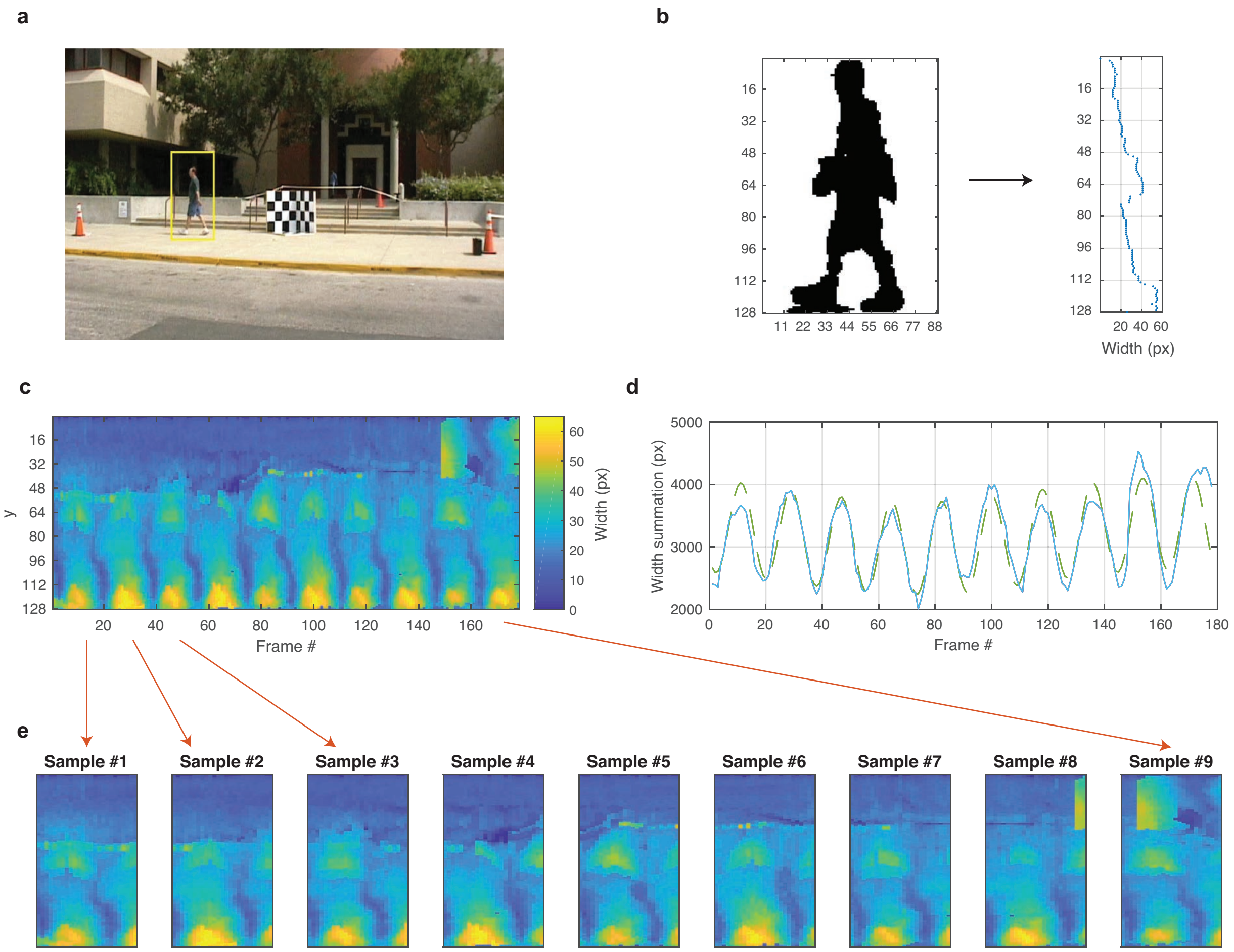

**Supplementary Figure 2: Pre-processing of the gait identification dataset. a,** One frame from the raw video. **b,** The extracted silhouette (ref. \cite{usf2002gait} in the main text) from the video, which was further converted to a width profile vector. Each dimension of the width profile vector represents the width of the silhouette at the corresponding height. **c,** The width profile vectors at each frame in the video. **d,** The total width in the width vector profile in each frame shows a periodic trend, which after processing a low-pass spectrum by an inverse Fourier transformation of the low-passed spectrum is used to detect the gait cycles. **e,** One video is divided into multiple samples according to the gait cycles.

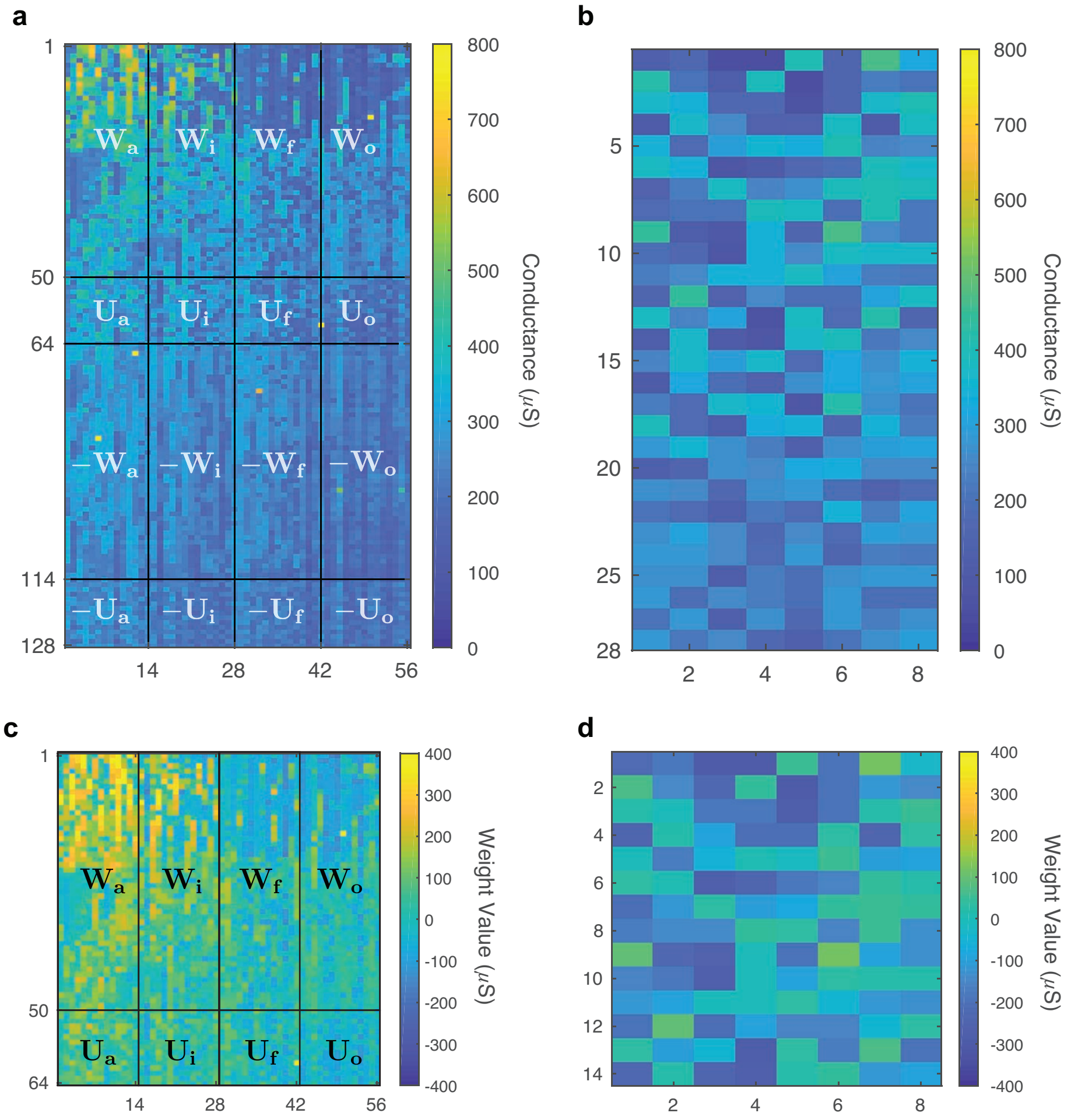

**Supplementary Figure 3: Additional data for the classification experiment. a,** Conductance map of the 128×56 memristors in the LSTM layer after the in-situ training. **b,** Conductance map of the 28×8 fully-connected layer after training. **c,** Map of synaptic weights calculated from the conductances shown in *(a)*. The LSTM synaptic weights are constituted by the weights ($\mathbf{W_a}$, $\mathbf{W_i}$, $\mathbf{W_f}$ and $\mathbf{W_o}$) connected to the input, and the recurrent weights ($\mathbf{U_a}$, $\mathbf{U_i}$, $\mathbf{U_f}$ and $\mathbf{U_o}$) connected to the LSTM outputs from the previous time step **d,** Map of the synaptic weights that calculated from the conductances shown in *(b)*.

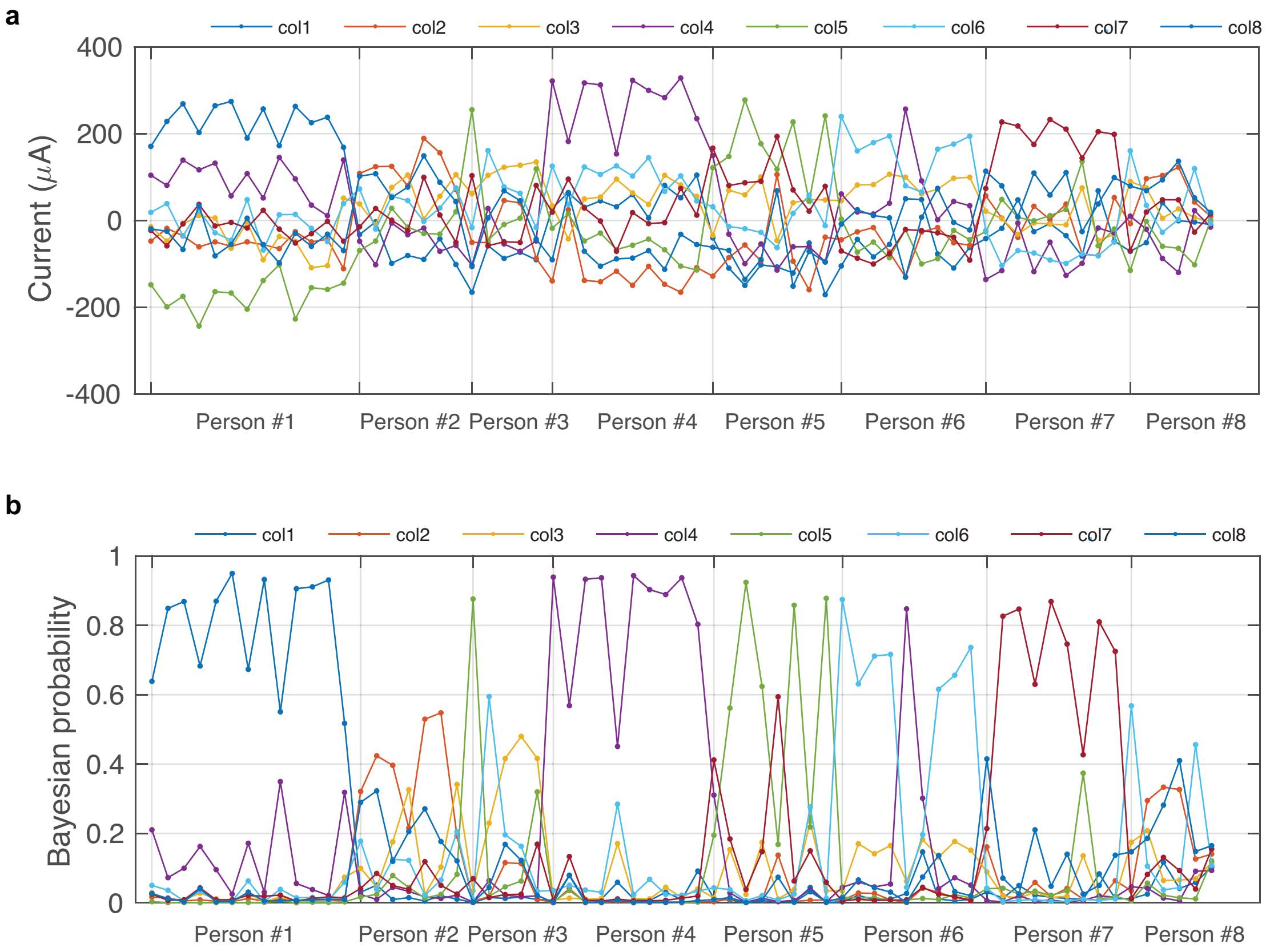

**Supplementary Figure 4: Output from the human identification inference. a,** Raw electrical current output after the in-situ training. Different curves represent the current output from different columns (col1 to col8). The maximum current output is identified as the inference result of the memristor RNN. **b,** The Bayesian probability computed from the data in *(a)* by the softmax function.